\documentclass[usenatbib,useAMS]{mn2e}
\usepackage{amssymb,amsmath,hyperref,graphics,graphicx}
\def\msol{M$_{\sun}$} \def\mdot{M$_{\sun}$\thinspace Myr$^{-1}$} \def\kms{km\thinspace s$^{-1}$}
\def\aap{A\&A}  \def\aj{AJ} \def\apj{ApJ} \def\apjl{ApJL} \def\apjs{ApJS} \def\araa{ARA\&A}  \def\mnras{MNRAS}
\def\memsai{Mem. Soc. Astron. Ital.} \def\nat{Nature}  \def\pasp{PASP}
\title[Multiple Populations in Globular Clusters]{The Celestial Buffet: multiple populations and globular cluster formation in dwarf galaxies}
\author[Maxwell, Wadsley, Couchman, Sills] {Aaron J. Maxwell\thanks{Email: ajmax@mcmaster.ca}, James Wadsley, H.M.P. Couchman, and Alison Sills\\
Department of Physics and Astronomy, McMaster University, Hamilton, Ontario L8S 4M1, Canada}
\date{Accepted 2014 January 13. Received 2014 January 7; in original form 2013 October 31.}
\begin{document}
\maketitle
\begin{abstract}
We present a framework that explains the commonly observed variation in light element abundances in globular clusters.
If globular clusters form in the centres of dwarf galaxies, they will be pumped on to larger orbits as star formation progresses.
The potential well will only retain the moderate velocity asymptotic giant branch (AGB) ejecta, the expected source of enrichment, but not supernova ejecta.
There is no need to increase the initial cluster mass, a requirement of self-enrichment scenarios, as all the stars within the dwarf can contribute.
As the clusters move through the dwarf centre they sweep up a mix of AGB ejecta and in-falling pristine gas to form a second generation of stars.
The specific mix will vary in time and is thus able to explain the spread in second generation abundances observed in different clusters.
The globular clusters will survive to the present day or be stripped as part of the hierarchical merging process of larger galaxies.
We illustrate how this process may operate using a high-resolution simulation of a dwarf galaxy at high redshift.
\end{abstract}
\section{Introduction}
Until recently, the standard picture of a globular cluster was that of a simple stellar population.
It was thought that all the stars formed in one time, in one place, and from a single cloud with a uniform chemical abundance.
However, high-precision photometry from the \emph{Hubble Space Telescope} \citep[e.g.][]{bedin2004, dantona2005, piotto2005, piotto2007, piotto2012} revealed split main sequences and sub-giant branches in many massive clusters, such as NGC 2808, M22, 47 Tuc, and NGC 1851.
Simultaneously, high-resolution spectroscopic studies of globular cluster stars \citep[e.g.][]{ramirez2001, carretta2009c} showed that almost all globular clusters have no star-to-star variations in iron abundance.
The variation in lighter elements, however, which had been characterized in bright giants for decades \citep[e.g.][]{carretta1994, cohen1999, gratton2006, carretta2007a}, was shown to extend down to stars on the main sequence \citep[e.g.][]{gratton2001, ramirez2002, carretta2003, carretta2004}.
Most surprisingly, a high He content is required to explain some of the observed properties of several of these clusters \citep[e.g.][]{piotto2005, dantona2005, carretta2007b, piotto2012}.\\
\indent The ubiquity of this light element spread in all well-studied globular clusters \citep*[see the review by][]{gratton2004} suggests globular clusters undergo a more complex formation process than that of a single burst of star formation.
The chemical patterns, the split photometric sequences, and extended horizontal branches \citep[e.g.][]{bedin2004} of globular clusters can be explained if, within the first few hundred million years of a cluster's existence, two or more populations of stars were formed \citep[e.g.][]{ventura2001, dantona2008}.
One was made from the same composition -- which we will call pristine throughout this paper -- as the stars of the halo: [Fe/H]\thinspace$\simeq$\thinspace-2, and $\alpha$-enriched but otherwise having scaled-solar abundances of the light elements.
The other generations were formed from material which has undergone hot hydrogen burning, occurring at temperatures above about $7\times10^{7}$\thinspace K.
Burning hydrogen under these circumstances uses the Ne--Na and Al--Mg cycles (analogues to the lower temperature C--N--O cycle), and can produce He as well as the observed trends of the other light elements.
The material which formed both populations, however, usually has the same Fe content.\\
\indent The commonly accepted explanation for these abundances is a sequence of events that a nascent globular cluster must undergo \citep[e.g.][]{dantona2008, ventura2008b}.
First, the proto-globular cluster forms from pristine gas.
After some time, the most massive stars explode as supernovae, but their ejecta have sufficient velocity to escape unhindered from the potential well of the cluster.
Later, a population of \emph{polluting} stars ejects their hot hydrogen-burnt material into the cluster at much lower velocities, so that the material is retained by the cluster.
Shortly thereafter, the second population forms from a mixture of this material and additional pristine material that has fallen into the cluster, in order to reproduce the observed abundance variations \citep[e.g.][]{carretta2009a, ventura2013}.
These two populations then passively evolve to become the present-day cluster.
This scenario broadly matches the observational constraints on the problem of multiple populations and extended horizontal branches in globulars, but there are two problems which we describe below.\\
\indent The majority of recent papers currently favour asymptotic giant branch (AGB) stars as the polluters \citep[e.g.][]{ventura2005b, dercole2010} since the bottom of their convective envelopes can produce the required overabundance of Na and N versus O and C \citep[e.g.][]{denissenkov1989}.
However, the nucleo synthetic yields from AGB stars need to be carefully tuned \citep[e.g.][]{denissenkov2003, james2004}, and there is still some uncertainty in the AGB evolution models \citep[e.g.][]{ventura2005a, ventura2008a}.
Due to these problems, other polluters have been proposed: rapidly rotating massive stars \citep[e.g.][]{decressin2007}, massive binary stars \citep[e.g.][]{demink2009}, and even stellar collisions \citep{sills2010}.\\
\indent The second problem has to do with the mass budget for the polluted population, which can make up to 50 per cent of the present cluster mass \citep[e.g.][]{carretta2009b, piotto2012}.
If one assumes a normal initial mass function (IMF) and an initial cluster mass that is close to its present-day mass ($\sim$\thinspace 10$^{6}$\thinspace\msol), then the population of polluting stars can only produce at most a few percent of the cluster mass as material with which to form the polluted population \citep[e.g.][]{cohen2005b}.
Most papers to date have addressed this issue by requiring the proto-cluster be at least 10 times more massive than the present cluster \citep[e.g.][]{dantona2008, ventura2008a, dercole2010, vesperini2013}, added more enriched gas to the ejecta by flattening the AGB range of the IMF \citep[e.g.][]{dantona2004, dantona2005}, or both.
Furthermore, highly unlikely star formation efficiencies of 100 per cent are required in the formation of the second population, or the mass-budget problem becomes even worse.\\
\indent \citet{bekki2010, bekki2011} simulated the formation of a second generation of stars from AGB ejecta within a cluster.
As expected \citep[e.g.][]{dantona2004, dercole2008}, a second population formed soon after the first starburst, but with a spatial and kinematic distribution completely different from the first population.
The simulations showed that the second population would be centrally condensed and show considerable rotation due to the dissipative processes required to drive the enriched material to realistic star-forming conditions. 
The initial cluster mass required to retain the AGB ejecta exceeded 6$\times10^{5}$\thinspace\msol, and even the best case scenario was only able to form 4$\times10^{4}$\thinspace\msol~in second population stars.
Yet, an order of magnitude increase in initial stellar mass would only produce enough AGB ejecta to form the \emph{present-day} mass of second generation stars.
Self-enrichment thus requires the cluster to be tidally stripped on time-scales much shorter than their relaxation times \citep[e.g.][]{vesperini2013}, since the first generation stars would have distributions initially extending to larger radii.\\
\indent The most straightforward solution to this problem was suggested by \citet{bekki2006b}: instead of treating the formation of globular clusters as simple stellar populations condensing from homogenous isolated gas clouds, they were treated as forming within the centres of dwarf galaxies at high redshift ($z$\thinspace$\gtrsim$\thinspace4).
This alleviated the mass budget problem, since now AGB ejecta from the surrounding dwarf galaxy spheroid would cool and settle to the centre, mix with the pristine material, and form the second population in the newly formed globular cluster.
However, even this scenario failed to reproduce the observed trends \citep{bekki2007}.\\
\indent The problem is that the simple approach of \citet{bekki2006b} would not make up the majority of globular clusters with variance only in the light elements.
\citet{bekki2006b} assumes the stellar nucleus of a dwarf progenitor is accreted on to a Milky Way (MW) sized halo, observable as a halo globular cluster.
However, the likelihood that SNe ejecta will be retained by the dwarf increases as its halo mass grows, imposing a limit on how long the stellar nucleus can be considered uniform in abundance.
This is evident in the broad range of Fe-enrichment exhibited by many of the Local Group Dwarfs, such as Fornax \citep[e.g.][]{pont2004}.
Many globular clusters show very little dispersion in the Fe-peak elements \citep[e.g.][]{ramirez2001, carretta2009c}, which suggests at least two possible constraints not discussed in \citet{bekki2006b}.
Either all globular clusters formed in dwarf progenitors that were accreted by larger haloes extremely early, or some process halted star formation in the nucleus on long time-scales, preserving the uniform iron abundances.\\
\indent There do exist peculiar globular clusters with dispersion in their heavy elements which would fit this model \citep[][]{bekki2006a}.
$\omega$ Cen is one example of a globular cluster with variations in [Fe/H] \citep[e.g.][]{norris1995, piotto2005}, which can be explained if it is the remnant stellar nucleus of an accreted dwarf galaxy \citep[e.g.][]{gnedin2002}.
Another is M54, located at the centre of the Sagittarius dwarf galaxy \citep[][]{ibata1995} and likely an example of $\omega$ Cen in an earlier accretion phase \citep[][]{carretta2010}.
NGC 2419 has similarly been argued to be the core of a stripped dwarf galaxy \citep{mackey2005, cohen2010, cohen2011, cohen2012}.
Clearly, what is lacking in the \citet{bekki2006b} model is a clear understanding of how the uniform heavy element abundance is preserved, if dwarf progenitors are the true sites of globular cluster formation.\\
\indent In this paper, we provide a new framework in which we can understand the formation of all globular clusters that exhibit abundance variations.
Like \citet{bekki2006b}, this new framework assumes the site of globular cluster formation is within the centres of dwarf galaxies.
\emph{Unlike} previous work, our framework proposes that these clusters are removed from the dwarf centres through dynamical evolution and end up on wide orbits, like those of the Fornax dwarf \citep[e.g.][]{hodge1961, mateo1998, letarte2006}, where they may be easily stripped.
By proposing a physically motivated mechanism for globular cluster removal, our new framework provides a consistent solution to the problem of abundance spreads with the cluster and links the probability of a spread in [Fe/H] to the amount of time spent in the dwarf centre.
We describe our new framework in \S\ref{sec:framework}, and provide an illustration of it in \S\ref{sec:illustrate} using a highly resolved simulation of a dwarf galaxy at high redshift \citep{mashchenko2008}.
We describe the setup in \S\ref{sub:setup}, with results in \S\ref{sub:results}.
\section{A New Framework For Forming Multiple Populations in Dwarf Galaxy Globular Clusters}\label{sec:framework}
The framework that will be outlined here rests on one key assumption: \textit{all globular clusters exhibiting abundance spreads formed near the centre of high-redshift dwarf galaxy progenitors and were later accreted during the hierarchical build-up of present-day massive galaxy haloes}.
As gas accretes on to the dwarf galaxy progenitor, it cools and collapses to the centre.
Once the gas reaches sufficient densities ($\gtrsim$\thinspace100 m$_{\text{H}}$\thinspace cm$^{-3}$) to form molecular clouds, star formation begins.
This will lead to feedback from massive stars in the form of radiation, winds, and supernova explosions that suppress star formation for about 30\thinspace Myr.
Massive amounts of gas will be swept out of the central regions, carrying the now $\alpha$-enriched material.\\
\indent The feedback will cause significant gas mass re-distribution within the dwarf galaxy centre, in turn altering the central gravitational potential.
Rapid fluctuations in the potential will lead to gravitational pumping of the collisionless components -- dark matter \citep{mashchenko2008, governato2012, pontzen2012} and stars \citep{maxwell2012, teyssier2013}.
A globular cluster formed within the centre will be moved to larger and larger orbits with each star formation burst.
The important point here is that even though the globular cluster is removed from the centre of the dwarf, \emph{it will make multiple passages through the gas-rich centre}.
On each pass, the globular cluster may accrete gas from the centre of the dwarf galaxy, including pollutants responsible for the light element abundance spread.
However, it will also experience subsequent energy kicks, eventually placed on so large an orbit that further accretion will be halted.\\
\indent Many groups have established through numerical experiments this energetic re-distribution of mass in dwarfs \citep[e.g.][]{read2005, mashchenko2006, mashchenko2008, governato2010, governato2012, pontzen2012}.
These studies focused on the transformation of the inner dark matter density profile from the cusps predicted by theory \citep[e.g.][]{dubinski1991, navarro1995, bullock2001, klypin2001, stadel2009} to the cores observed in Local Group dwarfs \citep[e.g.][]{burkert1995, cote2000, gilmore2007, oh2011}.
It has only been recently that attention has been paid to how this process would affect stars \citep{maxwell2012, teyssier2013}.
\citet{maxwell2012} focused on how this process would form spheroidal light profiles in the old stellar population, and by extension the presence of globular clusters at large projected radii from their hosts.
In our framework, both the formation of dark matter cores and multiple population globular clusters are intimately linked through the same mechanism of mass re-distribution.\\
\indent The centre of dwarf galaxies also contains a much deeper potential well than that of an isolated gas cloud or globular cluster.
Assuming that AGB stars are in fact the polluters, their wind can be retained within the dwarf nuclei \citep[e.g.][]{bekki2006a} since the speed at which the wind travels from the stellar surface is about 40\thinspace\kms~\citep[e.g.][]{woitke2006}.
Supernovae can blow out gas at upwards of 500\thinspace\kms~which can easily escape dwarf galaxies.
Globular clusters have typical escape speeds \citep{harris1996, gnedin2002} of 10--20\thinspace\kms~and so would be unable to retain this hot gas.
On the other hand, the AGB ejecta is retained in the deep potential and available for accretion \citep[e.g.][]{conroy2011} by globular clusters as they pass through the centre of the dwarf.
The clusters do not need to begin with masses an order of magnitude greater than that presently observed to cause self-inflicted pollution \citep[e.g.][]{cottrell1981}; instead, they draw from a reservoir created by the surrounding stars, provided the gas is accreted efficiently.\\
\indent Current models of the formation of the second population require some sort of dilution \citep[e.g.][]{carretta2009a} of the polluted material with pristine gas in order to create the observed abundance anticorrelations.
Since our framework places the formation site of the mixed abundance clusters within progenitor dwarf galaxies at high redshift, there should be plenty of gas in fall to lend itself to dilution \citep[e.g.][]{maxwell2012}.
Eventually, the gas within the centre will become predominantly pristine and the cluster formation process can begin anew.
A single dwarf galaxy could make several mixed abundance globular clusters within a few hundred Myr, long before Type Ia SNe begin to enrich the gas with Fe.
This is in sharp contrast to the work of \citet{bekki2006b} which would be more suitable for producing the more unusual objects that show clear [Fe/H] variations, such as $\omega$ Cen.
\section{An Illustration}\label{sec:illustrate}
We use the cosmological simulation of a well resolved dwarf galaxy by \citet{mashchenko2008} to demonstrate the salient points of our framework.
This simulation has been extensively studied in the context of the cusp--core problem \citep[for a recent review see][]{deblok2010} and the formation of Fornax-like spheroidal systems \citep{maxwell2012}.
The 12\thinspace pc force softening used in the simulation is comparable in scale to globular clusters and molecular clouds hosting star formation, but is still adequate for our purposes.
Within the simulation, \citet{maxwell2012} identified four bound star clusters over 100 times denser than the surrounding stellar spheroid that could be traced over 100\thinspace Myr.\\
\indent However, the resolution was not high enough to resolve the internal structure of the clusters, and so we cannot measure dynamical properties such as their mass distribution or velocity dispersion.
At these scales, accurate treatment of the formation of stars and the resultant feedback is required, which prevents us from directly studying the accretion of gas on to the cluster and the true mass of the clusters themselves.
Since our framework applies to any collisionless component of matter, we need only use a suitable globular cluster tracer throughout the simulation to illustrate it in a cosmological context.
Therefore, in the following setup, we use only the orbital properties of these clusters and treat the cluster mass as a free parameter.
\subsection{Accretion}\label{sub:setup}
Since we cannot directly measure accretion on to a star cluster as it passes through the gas-rich centre of the dwarf, we use the first-order estimate of \citet{bondi1944}:
\begin{equation}\label{eq:bhacc}
\dot{M}\simeq2\alpha\upi\frac{G^{2}M^{2}}{(v_{\text{rel}}^{2}+c_{\text{s}}^{2})^{3/2}}\bar{\rho},
\end{equation}
where $M$ is the mass of the cluster, $\bar{\rho}$ is the ambient gas density, $v_{\text{rel}}$ is the relative velocity between the cluster and the gas, and $c_{\rm{s}}$ is the sound speed of the gas.
The numerical factor $\alpha$ lies between 1 and 2 for most cases \citep{bondi1944,bondi1952}, but we have assumed unity so that we may be conservative in our estimate of the accretion rate.
Since we cannot directly measure the local gas density and temperature, we average the gas particle properties over a 35\thinspace pc sphere around the centre of mass of each cluster in each simulation snapshot.
This is the typical tidal radius for the MW clusters \citep{harris1996} derived from the King surface density profiles \citep{king1962, king1966}, and similar to the maximum accretion radius derived from Equation \eqref{eq:bhacc} for a 10$^{6}$\thinspace\msol~cluster and a sound speed of 10\thinspace\kms.
To compute the bulk relative velocity, we use the mass-weighted relative velocity with respect to the centre of mass velocity of the cluster:
\begin{equation}\label{eq:v}
\vec{v}_{\text{rel}}=\frac{\sum{m_{i}(\vec{v}_{i}-\vec{v}_{\text{com}})}}{\sum{m_{i}}},
\end{equation}
for all gas particles within the 35\thinspace pc sphere whose temperature is below $1.5\times10^{4}$\thinspace K.\\
\indent The original derivation of Equation \eqref{eq:bhacc} was for spherically symmetric accretion of a point mass moving through a uniform medium whose properties were measured very far from the point mass.
\citet{lin2007} have shown that for extended mass distributions whether Equation \eqref{eq:bhacc} applies to the cluster as a whole, or to individual stars within the cluster, depends on the internal velocity distribution of the stars.
Although we cannot directly measure the velocity distribution of the stars within the four clusters, the functional form of the accretion rate is preserved in both scenarios \citep{lin2007}.
Any uncertainty will be contained mainly in $\alpha$, which requires detailed numerical study \citep[e.g.][]{naiman2011}.
Since each of the four clusters spends significant time with relative speeds of 20--30\thinspace\kms~with respect to the surrounding gas, and given the spherical symmetry of globular clusters, Equation \eqref{eq:bhacc} should give a good estimate of the amount of gas accreted by a globular cluster moving through regions of dense gas \citep{conroy2011}.\\
\indent Once the gas has accreted on to the `surface' of a globular cluster, it should disperse throughout the cluster on a very short time-scale.
Using a mean half-mass radius of 4.3\thinspace pc \citep{harris1996} and a typical sound speed of 10\thinspace\kms~yields a crossing time of 0.4\thinspace Myr.
This is significantly shorter than the cooling time for the accreted gas and the onset of star formation, which is expected to last 2--3\thinspace Myr \citep[e.g.][]{dercole2008, bekki2011}, which is still shorter than the 10--20\thinspace Myr length of a typical accretion event experienced by the four clusters.
Thus, once gas is accreted it will quickly condense to the centre of the cluster and begin to form stars.\\
\subsubsection{Pollution Source}
We will assume that AGB stars are the source of the pollutants responsible for the light element abundance dispersion \citep[e.g.][]{denissenkov1989, dantona2008}, and that the winds from these stars distribute the pollutants.
However, our framework is not tied to a specific polluter and so will be applicable regardless of whether AGB stars are the true culprit; all that we require is that the source is present within the dwarf galaxy.
Most of the stars within the dwarf galaxy are found within 1\thinspace kpc \citep{maxwell2012} and the escape velocity from this radius is 60\thinspace\kms, so we can safely assume that the AGB wind will stay bound to the galaxy.\\
\indent Recently, \citet{larsen2012} suggested that the star formation history of Fornax placed severe constraints on the AGB mass available.
Although the simulation of \citet{mashchenko2008} did not track the light element abundances of individual gas particles due to AGB feedback, we can verify that the star formation history of the dwarf galaxy would satisfy even the highest observed fraction of second generation -- in other words, polluted -- stars by mass.
Since the star particles formed in the \citet{mashchenko2008} simulation represent many stars, we must integrate over the IMF to obtain the fraction of each star particle that would be expected to contribute to enriching the surrounding gas.
Given the uncertainties in AGB yields, we will focus only on the 3--6\thinspace\msol~mass range \citep[e.g.][]{ventura2001}, although 6--8\thinspace\msol~stars may also be a contributor \citep[e.g.][]{dercole2012}.
Using a typical power-law index $\alpha=-2.3$ \citep{salpeter1955, miller1979, kroupa2001, chabrier2003} over the mass range 0.1--100\thinspace\msol, approximately 8\thinspace\msol~per 100\thinspace\msol~will undergo the AGB phase; increasing the upper limit to 8\thinspace\msol~would add roughly an extra 3\thinspace\msol~per 100\thinspace\msol.
Assuming AGB stars lose at least 10 per cent of their initial mass over a period of 30--100\thinspace Myr yields a mean wind-loss rate of 10$^{-2}$\thinspace\mdot.
Converting the star formation history of the dwarf into an AGB ejecta history yields over $10^{5}$\thinspace\msol~of pure AGB ejecta within 1\thinspace kpc over a few Myr.\\
\indent In order to determine if this satisfies the observational constraints, we searched the literature \citep{ramirez2002, ramirez2003, cohen2005a, cohen2005c, carretta2006, carretta2007a, carretta2007b, carretta2007c, carretta2009a, carretta2009b} for spectroscopic measurements of the Na--O anti correlation, and follow \citet{carretta2009a} by splitting the stars into three components.
We then used their simple dilution model to estimate that $\sim$\thinspace7 per cent of the accreted mass needs to be composed of pure AGB ejecta in order to reproduce the global Na--O anticorrelation. 
In other words, a cluster whose final mass is $4\times10^{6}$\thinspace\msol~cluster with half of the stars showing signatures of enrichment would only require $1.5\times10^{5}$\thinspace\msol~of AGB ejecta.
Furthermore, the diffusion time of the AGB ejecta through the inner 1\thinspace kpc of the dwarf galaxy is
\begin{equation*}
t_{diff}\sim\frac{1\text{\thinspace kpc}}{v_{wind}}\simeq24\text{\thinspace Myr}.
\end{equation*}
This suggests that there may be inhomogeneity in the amount of enrichment within the gas pool from which a cluster may accrete, further diversifying the amount of dispersion a given cluster will exhibit.
\subsection{Results}\label{sub:results}
\begin{figure}
\centering
\includegraphics[width=8.5cm]{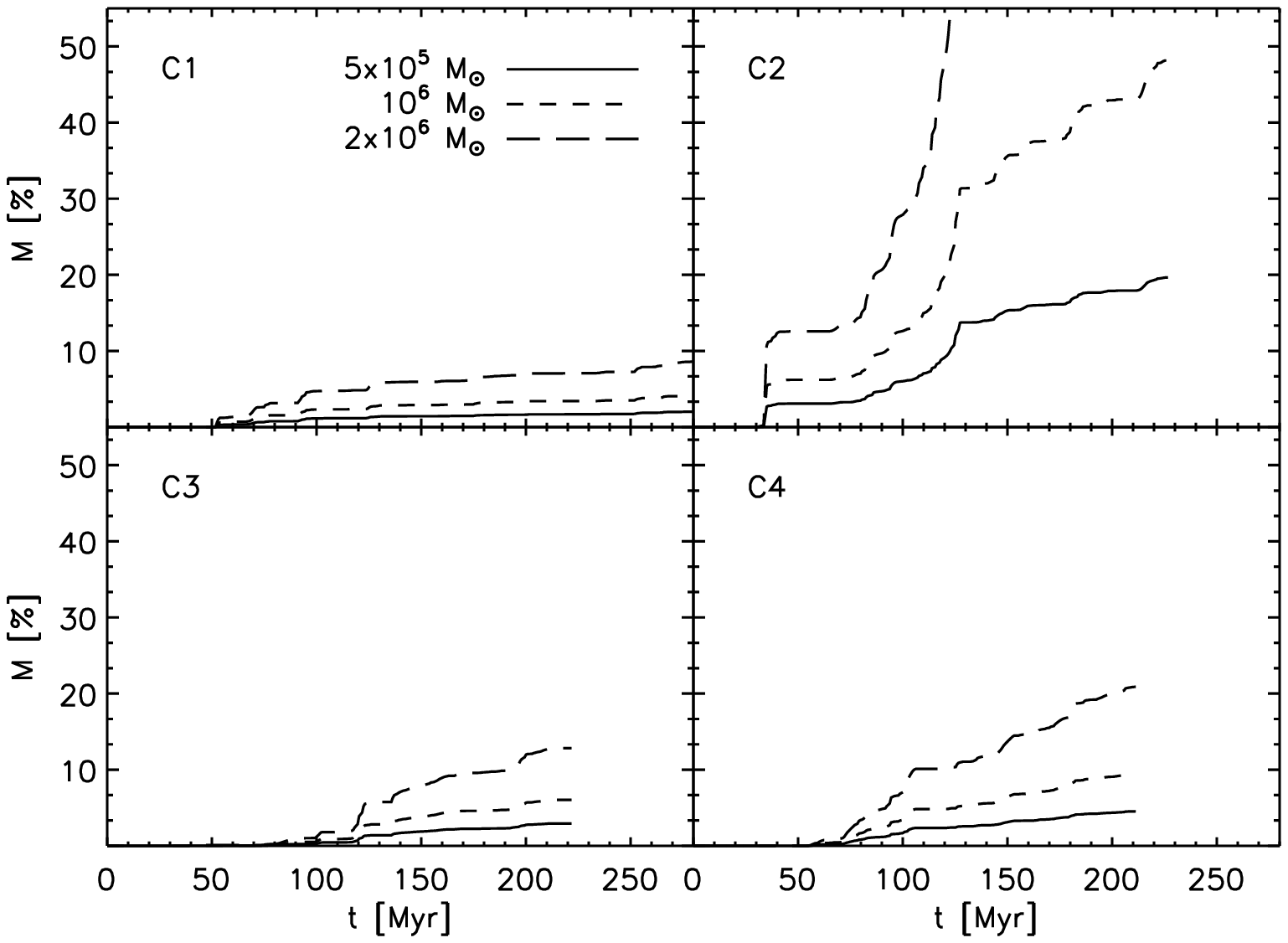}
\caption[fig1]{The percentage increase in mass estimated from Equation \eqref{eq:bhacc} as a function of time for the four clusters within the \citet{mashchenko2008} simulation.
The three lines represent the three different initial masses: $5\times10^{5}$\thinspace\msol~(solid), $10^{6}$\thinspace\msol~(short dashed), and $2\times10^{6}$\thinspace\msol~(long dashed).
The fraction by which a cluster can increase its mass depends on the varying orbit, initial cluster mass, and gas density within the Bondi--Hoyle radius.}
\label{fig:1}
\end{figure}

\begin{figure}
\centering
\includegraphics[width=8.5cm]{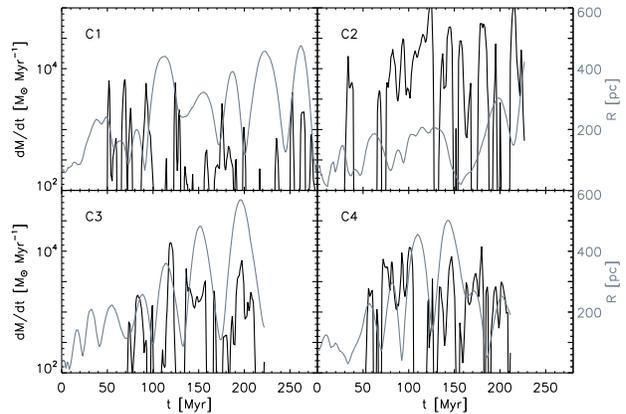} 
\caption[fig2]{The accretion rate for each of the four clusters predicted from Equation \eqref{eq:bhacc} using an initial mass of 2$\times10^{6}$\thinspace\msol, shown in black.
We have applied a boxcar filter to the accretion rate to remove noise caused by the time-dependent sampling.
We have also plotted the cluster orbital radius in grey.}
\label{fig:2}
\end{figure}

\begin{figure}
\centering
\includegraphics[width=8.5cm]{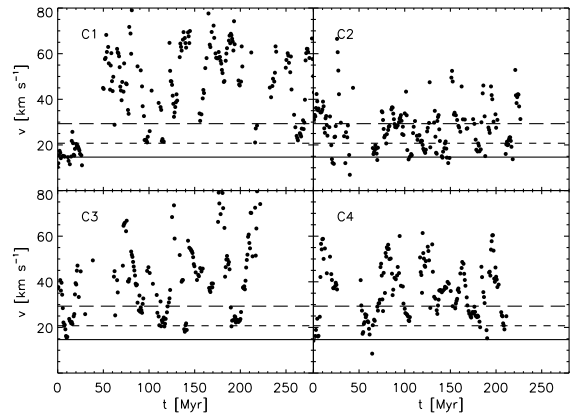} 
\caption[fig3]{The filled circles represent the temporal behaviour of $v_{rel}$.
The three solid lines represent escape velocities from 10\thinspace pc for each of the three cluster masses used in Fig. \ref{fig:1}.}
\label{fig:3}
\end{figure}

In order to find the potential mass growth of the four clusters traced within the simulation, we numerically integrate Equation \eqref{eq:bhacc}:
\begin{equation}\label{eq:growth}
M(t)=\int_{t_{o}}^{t}{\dot{M}\text{d}t'}.
\end{equation}
We start the integration 30\thinspace Myr after the formation of each cluster since this represents the end of the SNe phase which will sweep out any residual gas from the formation of the initial stellar population.
This allows sufficient time for the gas that formed the first generation cluster stars to be swept away by Type II supernovae.
This is supported by the observation that the majority of the star formation within the simulation occurs in bursts separated by 50--100\thinspace Myr \citep{maxwell2012}.\\
\indent Since the mass growth is highly non-linear, we will represent it as the percentage increase in mass as a function of time:
\begin{equation}\label{eq:mass}
\frac{M(t)-M(t_{o})}{M(t_{o})}\times\text{per cent},
\end{equation}
where $M(t_{o})$ corresponds to the initial cluster mass.
This is shown in Fig. \ref{fig:1}, for three different initial masses: $5\times10^{5}$\thinspace\msol~as the solid line, $10^{6}$\thinspace\msol~as the short dashed line, and $2\times10^{6}$\thinspace\msol~as the long dashed line.
The abscissa has been set to start at the formation time of each cluster.
Each cluster experiences wildly different growth rates, despite living in the same dynamic halo.\\
\indent First, the most massive clusters will accrete the most material at later times.
It has been observed that the strength of the Na--O anticorrelation in globular clusters is correlated with the cluster mass \citep[e.g.][]{recioblanco2006, carretta2009b, carretta2009c}.
Furthermore, the extent of the Na--O anticorrelation can be reproduced using a model wherein one source of material, either the pure AGB ejecta or the pristine gas, is diluted by the other.
In other words, there exist two time-scales: one for the accumulation of pristine material, and one for the accumulation of AGB ejecta.
Presumably, pristine material will accumulate at a rate dependent on the dwarf galaxy merger history, whereas the AGB ejecta will accumulate depending on the star formation history.
If the most massive clusters can accrete more gas for a longer time during each pass through the centre, then our framework applied to the \citet{mashchenko2008} simulation suggests that the pristine gas accumulates first, so that the more massive clusters can accumulate more AGB ejecta later.\\
\indent Secondly, the orbit of a cluster through its host dwarf progenitor primarily determines its mass growth.
The orbits of the clusters, shown in grey in Fig. \ref{fig:2}, grow with time due to the fluctuations in the gravitational potential induced by the re-distribution of the central dark matter mass.
This is a purely stochastic process, since it depends on both the rate of gas accumulation in the dwarf progenitor centre, the star formation rate, and the supernovae rate.
Each cluster will have a unique accretion history, even within the same dwarf galaxy progenitor, due to the varying number of AGB stars and their location within the dwarf, as well as the changing orbit.
This is consistent with the observation that the amount of light element enrichment per MW globular cluster varies between 10 and 50 per cent by mass \citep[e.g.][]{piotto2012}.\\
\indent Thus, we can consider the ratio of gas density to relative gas velocity as the accretion efficiency; a massive cluster passing quickly through a dense gas region may experience an accretion rate much lower than that of a lower mass cluster passing slowly through sparse pockets of gas.
The mass increase experienced by a cluster is a discontinuous process: clusters experience `growth spurts' as they pass through the centre of the dwarf galaxy progenitor where the densest gas is found.
The split main sequences within the globular clusters \citep[e.g.][]{piotto2012} would arise over time through the gradual combination of enriched and pristine material \citep[e.g.][]{bedin2004, piotto2005, piotto2007}.
The mass growth cannot continue indefinitely, however, as each boost in the cluster's orbit means that its relative velocity through the dwarf progenitor centre will increase, as shown in Fig. \ref{fig:2}.
The black lines show the accretion rate given by Equation \eqref{eq:bhacc} as a function of time for an initial mass of $2\times10^{6}$\thinspace\msol.\\
\indent Increasing the relative velocity of a cluster through dense gas also increases the probability that the clusters may experience ram pressure stripping.
Although Equation \eqref{eq:bhacc} does not take this into account, we can use the temporal behaviour of $v_{rel}$ of each cluster through the dense gas, shown in Fig. \ref{fig:3}, to determine whether the accreted gas is susceptible to removal by hydrodynamic forces.
Stripping will occur for globular clusters when the pressure of the accreted gas is less than the ram pressure of the ISM as it flows past the cluster.
Ignoring the cooling and gravitational collapse of the accreted gas, this condition is satisfied when $v_{rel}\gtrsim v_{esc}$, the cluster escape velocity, for most situations \citep{mori2000}.
In Fig. \ref{fig:3}, this is represented by the three horizontal lines which correspond to the escape velocity from 10\thinspace~pc for each of the cluster masses used in Fig. \ref{fig:1}.
It is clear that each cluster spends a significant amount of time within 35\thinspace pc of gas with relative speeds of 20--30\thinspace\kms.\\
\indent In general, the initial orbit of the cluster will significantly affect the ability for enriched gas to be accreted.
The three clusters with the largest orbits would have accreted 10--20\% of their initial mass, despite making multiple passes through the inner 100\thinspace pc of the dwarf galaxy progenitor.
The cluster with the highest estimated mass growth accretes much of its material during the 100\thinspace Myr when its orbit is least eccentric.
Finally, it experiences a huge energy boost that ejects the cluster past 300\thinspace pc, and were the simulation continued, it would probably experience a cut-off similar to that exhibited by the other three clusters.\\
\section{Summary}
We have proposed a new framework for the formation of multiple populations in dwarf galaxies.
In this framework, the high-redshift progenitors of dwarf galaxies are the formation sites of globular clusters with light element abundance dispersions.
The deeper potential well of the dwarf progenitors can easily retain the winds from AGB stars, thought to be the most likely source of the polluting material.
Fluctuations in the gravitational potential, caused by the re-distribution of matter by star formation feedback ocurring at the centres of dwarf galaxies, will drive growth in the clusters orbit.
In time, it will make multiple passes through the gas-rich dwarf centre, accreting a combination of pristine and polluted material.\\
\indent We have examined this framework in the context of the first cosmological simulation of a highly resolved dwarf galaxy.
Our results suggest that this framework broadly matches the mounting observational evidence of multiple populations in many, if not all, globular clusters.
It suggests a timeline for enrichment that matches the dilution models used to explain the observed light element anticorrelations, such as that in Na--O, with the observation that more massive clusters have the largest abundance spreads.
It also connects the stochastic nature of star formation and feedback to the observed spread in the number of second generation stars within each cluster, which is between 10 and 50 per cent by mass.
Finally, it provides the clearest difference between our new framework and those previously proposed, since our framework provides the blueprint to form the whole population of globular clusters, not just those with heavy element abundance spreads.
Thus, there exists at least two modes of stellar cluster formation within dwarf galaxies: the globular cluster channel and the stripped stellar nucleus channel.
\section*{acknowledgements}
We thank our anonymous referee whose careful reading of this manuscript and insightful comments helped improve this paper.
AJM, JW, HMPC, and AS are supported by NSERC.
This work has made use of the SHARCNET Consortium, part of the Compute/Calcul Canada Network.
This research has made use of NASA's Astrophysics Data System, and of the NASA/IPAC Extragalactic Database (NED) which is operated by the Jet Propulsion Laboratory, California Institute of Technology, under contract with the National Aeronautics and Space Administration.
This research has made use of the VizieR catalogue access tool, CDS, Strasbourg, France.
\bibliographystyle{mn2e}

\begin{thebibliography}{96}
\expandafter\ifx\csname natexlab\endcsname\relax\def\natexlab#1{#1}\fi
\bibitem[{{Bedin} {et~al}\mbox{.}(2004){Bedin}, {Piotto}, {Anderson}, {Cassisi}, {King}, {Momany}, \& {Carraro}}]{bedin2004}
{Bedin} L.~R., {Piotto} G., {Anderson} J., {Cassisi} S., {King} I.~R., {Momany} Y., {Carraro} G., 2004, \apjl, 605, L125
\bibitem[{{Bekki}(2006)}]{bekki2006b}
{Bekki} K., 2006, \mnras, 367, L24
\bibitem[{{Bekki}(2010)}]{bekki2010}
{Bekki} K., 2010, \apj, 724, L99
\bibitem[{{Bekki}(2011)}]{bekki2011}
{Bekki} K., 2011, \mnras, 412, 2241
\bibitem[{{Bekki} {et~al}\mbox{.}(2007){Bekki}, {Campbell}, {Lattanzio}, \& {Norris}}]{bekki2007}
{Bekki} K., {Campbell} S.~W., {Lattanzio} J.~C., {Norris} J.~E., 2007, \mnras, 377, 335
\bibitem[{{Bekki} \& {Norris}(2006)}]{bekki2006a}
{Bekki} K., {Norris} J.~E., 2006, \apj, 637, L109
\bibitem[{{Bondi}(1952)}]{bondi1952}
{Bondi} H., 1952, \mnras, 112, 195
\bibitem[{{Bondi} \& {Hoyle}(1944)}]{bondi1944}
{Bondi} H., {Hoyle} F., 1944, \mnras, 104, 273
\bibitem[{{Bullock} {et~al}\mbox{.}(2001){Bullock}, {Kolatt}, {Sigad}, {Somerville}, {Kravtsov}, {Klypin}, {Primack}, \& {Dekel}}]{bullock2001}
{Bullock} J.~S., {Kolatt} T.~S., {Sigad} Y., {Somerville} R.~S., {Kravtsov} A.~V., {Klypin} A.~A., {Primack} J.~R., {Dekel} A., 2001, \mnras, 321, 559
\bibitem[{{Burkert}(1995)}]{burkert1995}
{Burkert} A., 1995, \apjl, 447, L25
\bibitem[{{Carretta} \& {Gratton}(1994)}]{carretta1994}
{Carretta} E., {Gratton} R., 1994, \memsai, 65, 799
\bibitem[{{Carretta} {et~al}\mbox{.}(2003){Carretta}, {Bragaglia}, {Cacciari},  \& {Rossetti}}]{carretta2003}
{Carretta} E., {Bragaglia} A., {Cacciari} C., {Rossetti} E., 2003, \aap, 410, 143
\bibitem[{{Carretta} {et~al}\mbox{.}(2004){Carretta}, {Gratton}, {Bragaglia}, {Bonifacio}, \& {Pasquini}}]{carretta2004}
{Carretta} E., {Gratton} R.~G., {Bragaglia} A., {Bonifacio} P., {Pasquini} L., 2004, \aap, 416, 925
\bibitem[{{Carretta} {et~al}\mbox{.}(2006){Carretta}, {Bragaglia}, {Gratton}, {Leone}, {Recio-Blanco}, \& {Lucatello}}]{carretta2006}
{Carretta} E., {Bragaglia} A., {Gratton} R.~G., {Leone} F., {Recio-Blanco} A., {Lucatello} S., 2006, \aap, 450, 523
\bibitem[{{Carretta} {et~al}\mbox{.}(2007{\natexlab{a}}){Carretta}, {Bragaglia}, {Gratton}, {Lucatello}, \& {Momany}}]{carretta2007a}
{Carretta} E., {Bragaglia} A., {Gratton} R.~G., {Lucatello} S., {Momany} Y., 2007{\natexlab{a}}, \aap, 464, 927
\bibitem[{{Carretta} {et~al}\mbox{.}(2007{\natexlab{b}}){Carretta}, {Bragaglia}, {Gratton}, {Catanzaro}, {Leone}, {Sabbi}, {Cassisi}, {Claudi}, {D'Antona}, {Fran{\c c}ois}, {James}, \& {Piotto}}]{carretta2007b}
{Carretta} E. {et~al.}, 2007{\natexlab{b}}, \aap, 464, 939
\bibitem[{{Carretta} {et~al}\mbox{.}(2007{\natexlab{c}}){Carretta}, {Bragaglia}, {Gratton}, {Momany}, {Recio-Blanco}, {Cassisi}, {Fran{\c c}ois}, {James}, {Lucatello}, \& {Moehler}}]{carretta2007c}
{Carretta} E. {et~al.}, 2007{\natexlab{c}}, \aap, 464, 967
\bibitem[{{Carretta} {et~al}\mbox{.}(2009{\natexlab{a}}){Carretta}, {Bragaglia}, {Gratton}, {Lucatello}, {Catanzaro}, {Leone}, {Bellazzini}, {Claudi}, {D'Orazi}, {Momany}, {Ortolani}, {Pancino}, {Piotto}, {Recio-Blanco}, \& {Sabbi}}]{carretta2009a}
{Carretta} E. {et~al.}, 2009{\natexlab{a}}, \aap, 505, 117
\bibitem[{{Carretta} {et~al}\mbox{.}(2009{\natexlab{b}}){Carretta}, {Bragaglia}, {Gratton}, \& {Lucatello}}]{carretta2009b}
{Carretta} E., {Bragaglia} A., {Gratton} R., {Lucatello} S., 2009{\natexlab{b}}, \aap, 505, 139
\bibitem[{{Carretta} {et~al}\mbox{.}(2009{\natexlab{c}}){Carretta}, {Bragaglia}, {Gratton}, {D'Orazi}, \& {Lucatello}}]{carretta2009c}
{Carretta} E., {Bragaglia} A., {Gratton} R., {D'Orazi} V., {Lucatello} S., 2009{\natexlab{c}}, \aap, 508, 695
\bibitem[{{Carretta} {et~al}\mbox{.}(2010){Carretta}, {Bragaglia}, {Gratton}, {Lucatello}, {Bellazzini}, {Catanzaro}, {Leone}, {Momany}, {Piotto}, \& {D'Orazi}}]{carretta2010}
{Carretta} E. {et~al.}, 2010, \apj, 714, L7
\bibitem[{{Chabrier}(2003)}]{chabrier2003}
{Chabrier} G., 2003, \pasp, 115, 763
\bibitem[{{Cohen}(1999)}]{cohen1999}
{Cohen} J.~G., 1999, \aj, 117, 2434
\bibitem[{{Cohen}, {Briley} \& {Stetson}(2005){Cohen}, {Briley}, \& {Stetson}}]{cohen2005b}
{Cohen} J.~G., {Briley} M.~M., {Stetson} P.~B., 2005, \aj, 130, 1177
\bibitem[{{Cohen}, {Huang} \& {Kirby}(2011){Cohen}, {Huang}, \& {Kirby}}]{cohen2011}
{Cohen} J.~G., {Huang} W., {Kirby} E.~N., 2011, \apj, 740, 60
\bibitem[{{Cohen} \& {Kirby}(2012)}]{cohen2012}
{Cohen} J.~G., {Kirby} E.~N., 2012, \apj, 760, 86
\bibitem[{{Cohen} {et~al}\mbox{.}(2010){Cohen}, {Kirby}, {Simon}, \& {Geha}}]{cohen2010}
{Cohen} J.~G., {Kirby} E.~N., {Simon} J.~D., {Geha} M., 2010, \apj, 725, 288
\bibitem[{{Cohen} \& {Mel{\'e}ndez}(2005{\natexlab{a}})}]{cohen2005a}
{Cohen} J.~G., {Mel{\'e}ndez} J., 2005{\natexlab{a}}, \aj, 129, 303
\bibitem[{{Cohen} \& {Mel{\'e}ndez}(2005{\natexlab{b}})}]{cohen2005c}
{Cohen} J.~G., {Mel{\'e}ndez} J., 2005{\natexlab{b}}, \aj, 129, 1607
\bibitem[{{Conroy} \& {Spergel}(2011)}]{conroy2011}
{Conroy} C., {Spergel} D.~N., 2011, \apj, 726, 36
\bibitem[{{C{\^o}t{\'e}}, {Carignan} \& {Freeman}(2000){C{\^o}t{\'e}}, {Carignan}, \& {Freeman}}]{cote2000}
{C{\^o}t{\'e}} S., {Carignan} C., {Freeman} K.~C., 2000, \aj, 120, 3027
\bibitem[{{Cottrell} \& {Da Costa}(1981)}]{cottrell1981}
{Cottrell} P.~L., {Da Costa} G.~S., 1981, \apj, 245, L79
\bibitem[{{D'Antona} {et~al}\mbox{.}(2005){D'Antona}, {Bellazzini}, {Caloi}, {Pecci}, {Galleti}, \& {Rood}}]{dantona2005}
{D'Antona} F., {Bellazzini} M., {Caloi} V., {Pecci} F.~F., {Galleti} S., {Rood} R.~T., 2005, \apj, 631, 868
\bibitem[{{D'Antona} \& {Caloi}(2004)}]{dantona2004}
{D'Antona} F., {Caloi} V., 2004, \apj, 611, 871
\bibitem[{{D'Antona} \& {Caloi}(2008)}]{dantona2008}
{D'Antona} F., {Caloi} V., 2008, \mnras, 390, 693
\bibitem[{{de Blok}(2010)}]{deblok2010}
{de Blok} W.~J.~G., 2010, Adv. Astron., 2010
\bibitem[{{de Mink} {et~al}\mbox{.}(2009){de Mink}, {Pols}, {Langer}, \& {Izzard}}]{demink2009}
{de Mink} S.~E., {Pols} O.~R., {Langer} N., {Izzard} R.~G., 2009, \aap, 507, L1
\bibitem[{{Decressin} {et~al}\mbox{.}(2007){Decressin}, {Meynet}, {Charbonnel}, {Prantzos}, \& {Ekstr{\"o}m}}]{decressin2007}
{Decressin} T., {Meynet} G., {Charbonnel} C., {Prantzos} N., {Ekstr{\"o}m} S., 2007, \aap, 464, 1029
\bibitem[{{Denisenkov} \& {Denisenkova}(1989)}]{denissenkov1989}
{Denisenkov} P.~A., {Denisenkova} S.~N., 1989, Astronomicheskij Tsirkulyar, 1538, 11
\bibitem[{{Denissenkov} \& {Herwig}(2003)}]{denissenkov2003}
{Denissenkov} P.~A., {Herwig} F., 2003, \apj, 590, L99
\bibitem[{{D'Ercole} {et~al}\mbox{.}(2012){D'Ercole}, {D'Antona}, {Carini}, {Vesperini}, \& {Ventura}}]{dercole2012}
{D'Ercole} A., {D'Antona} F., {Carini} R., {Vesperini} E., {Ventura} P., 2012, \mnras, 423, 1521
\bibitem[{{D'Ercole} {et~al}\mbox{.}(2010){D'Ercole}, {D'Antona}, {Ventura}, {Vesperini}, \& {McMillan}}]{dercole2010}
{D'Ercole} A., {D'Antona} F., {Ventura} P., {Vesperini} E., {McMillan} S.~L.~W., 2010, \mnras, 407, 854
\bibitem[{{D'Ercole} {et~al}\mbox{.}(2008){D'Ercole}, {Vesperini}, {D'Antona}, {McMillan}, \& {Recchi}}]{dercole2008}
{D'Ercole} A., {Vesperini} E., {D'Antona} F., {McMillan} S.~L.~W., {Recchi} S., 2008, \mnras, 391, 825
\bibitem[{{Dubinski} \& {Carlberg}(1991)}]{dubinski1991}
{Dubinski} J., {Carlberg} R.~G., 1991, \apj, 378, 496
\bibitem[{{Gilmore} {et~al}\mbox{.}(2007){Gilmore}, {Wilkinson}, {Wyse}, {Kleyna}, {Koch}, {Evans}, \& {Grebel}}]{gilmore2007}
{Gilmore} G., {Wilkinson} M.~I., {Wyse} R.~F.~G., {Kleyna} J.~T., {Koch} A., {Evans} N.~W., {Grebel} E.~K., 2007, \apj, 663, 948
\bibitem[{{Gnedin} {et~al}\mbox{.}(2002){Gnedin}, {Zhao}, {Pringle}, {Fall}, {Livio}, \& {Meylan}}]{gnedin2002}
{Gnedin} O.~Y., {Zhao} H., {Pringle} J.~E., {Fall} S.~M., {Livio} M., {Meylan} G., 2002, \apjl, 568, L23
\bibitem[{{Governato} {et~al}\mbox{.}(2010){Governato}, {Brook}, {Mayer}, {Brooks}, {Rhee}, {Wadsley}, {Jonsson}, {Willman}, {Stinson}, {Quinn}, \& {Madau}}]{governato2010}
{Governato} F. {et~al.}, 2010, \nat, 463, 203
\bibitem[{{Governato} {et~al}\mbox{.}(2012){Governato}, {Zolotov}, {Pontzen}, {Christensen}, {Oh}, {Brooks}, {Quinn}, {Shen}, \& {Wadsley}}]{governato2012}
{Governato} F. {et~al.}, 2012, \mnras, 422, 1231
\bibitem[{{Gratton}, {Sneden} \& {Carretta}(2004){Gratton}, {Sneden}, \& {Carretta}}]{gratton2004}
{Gratton} R., {Sneden} C., {Carretta} E., 2004, \araa, 42, 385
\bibitem[{{Gratton} {et~al}\mbox{.}(2001){Gratton}, {Bonifacio}, {Bragaglia}, {Carretta}, {Castellani}, {Centurion}, {Chieffi}, {Claudi}, {Clementini}, {D'Antona}, {Desidera}, {Fran{\c c}ois}, {Grundahl}, {Lucatello}, {Molaro}, {Pasquini}, {Sneden}, {Spite}, \& {Straniero}}]{gratton2001}
{Gratton} R.~G. {et~al.}, 2001, \aap, 369, 87
\bibitem[{{Gratton} {et~al}\mbox{.}(2006){Gratton}, {Lucatello}, {Bragaglia}, {Carretta}, {Momany}, {Pancino}, \& {Valenti}}]{gratton2006}
{Gratton} R.~G., {Lucatello} S., {Bragaglia} A., {Carretta} E., {Momany} Y., {Pancino} E., {Valenti} E., 2006, \aap, 455, 271
\bibitem[{{Harris}(1996)}]{harris1996}
{Harris} W.~E., 1996, \aj, 112, 1487
\bibitem[{{Hodge}(1961)}]{hodge1961}
{Hodge} P.~W., 1961, \aj, 66, 83
\bibitem[{{Ibata}, {Gilmore} \& {Irwin}(1995){Ibata}, {Gilmore}, \& {Irwin}}]{ibata1995}
{Ibata} R.~A., {Gilmore} G., {Irwin} M.~J., 1995, \mnras, 277, 781
\bibitem[{{James} {et~al}\mbox{.}(2004){James}, {Fran{\c c}ois}, {Bonifacio}, {Carretta}, {Gratton}, \& {Spite}}]{james2004}
{James} G., {Fran{\c c}ois} P., {Bonifacio} P., {Carretta} E., {Gratton} R.~G., {Spite} F., 2004, \aap, 427, 825
\bibitem[{{King}(1962)}]{king1962}
{King} I., 1962, \aj, 67, 471
\bibitem[{{King}(1966)}]{king1966}
{King} I.~R., 1966, \aj, 71, 64
\bibitem[{{Klypin} {et~al}\mbox{.}(2001){Klypin}, {Kravtsov}, {Bullock}, \& {Primack}}]{klypin2001}
{Klypin} A., {Kravtsov} A.~V., {Bullock} J.~S., {Primack} J.~R., 2001, \apj, 554, 903
\bibitem[{{Kroupa}(2001)}]{kroupa2001}
{Kroupa} P., 2001, \mnras, 322, 231
\bibitem[{{Larsen}, {Strader} \& {Brodie}(2012){Larsen}, {Strader}, \& {Brodie}}]{larsen2012}
{Larsen} S.~S., {Strader} J., {Brodie} J.~P., 2012, \aap, 544, L14
\bibitem[{{Letarte} {et~al}\mbox{.}(2006){Letarte}, {Hill}, {Jablonka}, {Tolstoy}, {Fran{\c c}ois}, \& {Meylan}}]{letarte2006}
{Letarte} B., {Hill} V., {Jablonka} P., {Tolstoy} E., {Fran{\c c}ois} P., {Meylan} G., 2006, \aap, 453, 547
\bibitem[{{Lin} \& {Murray}(2007)}]{lin2007}
{Lin} D.~N.~C., {Murray} S.~D., 2007, \apj, 661, 779
\bibitem[{{Mackey} \& {van den Bergh}(2005)}]{mackey2005} 
{Mackey} A.~D., {van den Bergh} S., 2005, \mnras, 360, 631
\bibitem[{{Mashchenko}, {Couchman} \& {Wadsley}(2006){Mashchenko}, {Couchman}, \& {Wadsley}}]{mashchenko2006}
{Mashchenko} S., {Couchman} H.~M.~P., {Wadsley} J., 2006, \nat, 442, 539
\bibitem[{{Mashchenko}, {Wadsley} \& {Couchman}(2008){Mashchenko}, {Wadsley}, \& {Couchman}}]{mashchenko2008}
{Mashchenko} S., {Wadsley} J., {Couchman} H.~M.~P., 2008, Science, 319, 174
\bibitem[{{Mateo}(1998)}]{mateo1998}
{Mateo} M.~L., 1998, \araa, 36, 435
\bibitem[{{Maxwell} {et~al}\mbox{.}(2012){Maxwell}, {Wadsley}, {Couchman}, \& {Mashchenko}}]{maxwell2012}
{Maxwell} A.~J., {Wadsley} J., {Couchman} H.~M.~P., {Mashchenko} S., 2012, \apj, 755, L35
\bibitem[{{Miller} \& {Scalo}(1979)}]{miller1979}
{Miller} G.~E., {Scalo} J.~M., 1979, \apjs, 41, 513
\bibitem[{{Mori} \& {Burkert}(2000)}]{mori2000}
{Mori} M., {Burkert} A., 2000, \apj, 538, 559
\bibitem[{{Naiman}, {Ramirez-Ruiz} \& {Lin}(2011){Naiman}, {Ramirez-Ruiz}, \& {Lin}}]{naiman2011}
{Naiman} J.~P., {Ramirez-Ruiz} E., {Lin} D.~N.~C., 2011, \apj, 735, 25
\bibitem[{{Navarro}, {Frenk} \& {White}(1995){Navarro}, {Frenk}, \& {White}}]{navarro1995}
{Navarro} J.~F., {Frenk} C.~S., {White} S.~D.~M., 1995, \mnras, 275, 720
\bibitem[{{Norris} \& {Da Costa}(1995)}]{norris1995}
{Norris} J.~E., {Da Costa} G.~S., 1995, \apj, 447, 680
\bibitem[{{Oh} {et~al}\mbox{.}(2011){Oh}, {de Blok}, {Brinks}, {Walter}, \& {Kennicutt}}]{oh2011}
{Oh} S.-H., {de Blok} W.~J.~G., {Brinks} E., {Walter} F., {Kennicutt}, R.~C., Jr, 2011, \aj, 141, 193
\bibitem[{{Piotto} {et~al}\mbox{.}(2007){Piotto}, {Bedin}, {Anderson}, {King}, {Cassisi}, {Milone}, {Villanova}, {Pietrinferni}, \& {Renzini}}]{piotto2007}
{Piotto} G. {et~al.}, 2007, \apj, 661, L53
\bibitem[{{Piotto} {et~al}\mbox{.}(2012){Piotto}, {Milone}, {Anderson}, {Bedin}, {Bellini}, {Cassisi}, {Marino}, {Aparicio}, \& {Nascimbeni}}]{piotto2012}
{Piotto} G. {et~al.}, 2012, \apj, 760, 39
\bibitem[{{Piotto} {et~al}\mbox{.}(2005){Piotto}, {Villanova}, {Bedin}, {Gratton}, {Cassisi}, {Momany}, {Recio-Blanco}, {Lucatello}, {Anderson}, {King}, {Pietrinferni}, \& {Carraro}}]{piotto2005}
{Piotto} G. {et~al.}, 2005, \apj, 621, 777
\bibitem[{{Pont} {et~al}\mbox{.}(2004){Pont}, {Zinn}, {Gallart}, {Hardy}, \& {Winnick}}]{pont2004}
{Pont} F., {Zinn} R., {Gallart} C., {Hardy} E., {Winnick} R., 2004, \aj, 127, 840
\bibitem[{{Pontzen} \& {Governato}(2012)}]{pontzen2012}
{Pontzen} A., {Governato} F., 2012, \mnras, 421, 3464
\bibitem[{{Ram{\'{\i}}rez} \& {Cohen}(2002)}]{ramirez2002}
{Ram{\'{\i}}rez} S.~V., {Cohen} J.~G., 2002, \aj, 123, 3277
\bibitem[{{Ram{\'{\i}}rez} \& {Cohen}(2003)}]{ramirez2003}
{Ram{\'{\i}}rez} S.~V., {Cohen} J.~G., 2003, \aj, 125, 224
\bibitem[{{Ram{\'{\i}}rez} {et~al}\mbox{.}(2001){Ram{\'{\i}}rez}, {Cohen}, {Buss}, \& {Briley}}]{ramirez2001}
{Ram{\'{\i}}rez} S.~V., {Cohen} J.~G., {Buss} J., {Briley} M.~M., 2001, \aj, 122, 1429
\bibitem[{{Read} \& {Gilmore}(2005)}]{read2005}
{Read} J.~I., {Gilmore} G., 2005, \mnras, 356, 107
\bibitem[{{Recio-Blanco} {et~al}\mbox{.}(2006){Recio-Blanco}, {Aparicio}, {Piotto}, {de Angeli}, \& {Djorgovski}}]{recioblanco2006}
{Recio-Blanco} A., {Aparicio} A., {Piotto} G., {de Angeli} F., {Djorgovski} S.~G., 2006, \aap, 452, 875
\bibitem[{{Salpeter}(1955)}]{salpeter1955}
{Salpeter} E.~E., 1955, \apj, 121, 161
\bibitem[{{Sills} \& {Glebbeek}(2010)}]{sills2010}
{Sills} A., {Glebbeek} E., 2010, \mnras, 407, 277
\bibitem[{{Stadel} {et~al}\mbox{.}(2009){Stadel}, {Potter}, {Moore}, {Diemand}, {Madau}, {Zemp}, {Kuhlen}, \& {Quilis}}]{stadel2009}
{Stadel} J., {Potter} D., {Moore} B., {Diemand} J., {Madau} P., {Zemp} M., {Kuhlen} M., {Quilis} V., 2009, \mnras, 398, L21
\bibitem[{{Teyssier} {et~al}\mbox{.}(2013){Teyssier}, {Pontzen}, {Dubois}, \& {Read}}]{teyssier2013}
{Teyssier} R., {Pontzen} A., {Dubois} Y., {Read} J.~I., 2013, \mnras, 429, 3068
\bibitem[{{Ventura} \& {D'Antona}(2005{\natexlab{a}})}]{ventura2005a}
{Ventura} P., {D'Antona} F., 2005{\natexlab{a}}, \aap, 439, 1075
\bibitem[{{Ventura} \& {D'Antona}(2005{\natexlab{b}})}]{ventura2005b}
{Ventura} P., {D'Antona} F., 2005{\natexlab{b}}, \apj, 635, L149
\bibitem[{{Ventura} \& {D'Antona}(2008{\natexlab{a}})}]{ventura2008b}
{Ventura} P., {D'Antona} F., 2008{\natexlab{a}}, \mnras, 385, 2034
\bibitem[{{Ventura} \& {D'Antona}(2008{\natexlab{b}})}]{ventura2008a}
{Ventura} P., {D'Antona} F., 2008{\natexlab{b}}, \aap, 479, 805
\bibitem[{{Ventura} {et~al}\mbox{.}(2001){Ventura}, {D'Antona}, {Mazzitelli}, \& {Gratton}}]{ventura2001}
{Ventura} P., {D'Antona} F., {Mazzitelli} I., {Gratton} R., 2001, \apj, 550, L65
\bibitem[{{Ventura} {et~al}\mbox{.}(2013){Ventura}, {Di Criscienzo}, {Carini}, \& {D'Antona}}]{ventura2013}
{Ventura} P., {Di Criscienzo} M., {Carini} R., {D'Antona} F., 2013, \mnras, 431, 3642
\bibitem[{{Vesperini} {et~al}\mbox{.}(2013){Vesperini}, {McMillan}, {D'Antona}, \& {D'Ercole}}]{vesperini2013}
{Vesperini} E., {McMillan} S.~L.~W., {D'Antona} F., {D'Ercole} A., 2013, \mnras, 429, 1913
\bibitem[{{Woitke}(2006)}]{woitke2006}
{Woitke} P., 2006, \aap, 452, 537
\end{thebibliography}

\end{document}